\documentclass[%
 aip,
 amsmath,amssymb,
 reprint,%
]{revtex4-1}

\usepackage{graphicx}
\usepackage{dcolumn}
\usepackage{bm}
\usepackage{comment}

\usepackage[utf8]{inputenc}
\usepackage[T1]{fontenc}
\usepackage{mathptmx}

\begin{document}


\title[A modular ultra-high vacuum millikelvin scanning tunneling microscope]{A modular ultra-high vacuum millikelvin scanning tunneling microscope}

\author{Dillon Wong}
    \thanks{These authors contributed equally to this work.}
    \affiliation{Joseph Henry Laboratories and Department of Physics, Princeton University, Princeton, New Jersey 08544, USA}
\author{Sangjun Jeon}
    \thanks{These authors contributed equally to this work.}
    \affiliation{Joseph Henry Laboratories and Department of Physics, Princeton University, Princeton, New Jersey 08544, USA}
    \affiliation{Department of Physics, Chung-Ang University, Seoul 06974, Republic of Korea}
\author{Kevin P. Nuckolls}
    \thanks{These authors contributed equally to this work.}
    \affiliation{Joseph Henry Laboratories and Department of Physics, Princeton University, Princeton, New Jersey 08544, USA}
\author{Myungchul Oh}
    \thanks{These authors contributed equally to this work.}
    \affiliation{Joseph Henry Laboratories and Department of Physics, Princeton University, Princeton, New Jersey 08544, USA}
\author{Simon C. J. Kingsley}
    \affiliation{Oxford Instruments, Tubney Woods, Abingdon, Oxfordshire, OX13 5QX,
United Kingdom}
\author{Ali Yazdani}
    \email{yazdani@princeton.edu}
    \affiliation{Joseph Henry Laboratories and Department of Physics, Princeton University, Princeton, New Jersey 08544, USA}

\date{\today}

\begin{abstract}
We describe the design, construction, and performance of an ultra-high vacuum (UHV) scanning tunneling microscope (STM) capable of imaging at dilution-refrigerator temperatures and equipped with a vector magnet. The primary objective of our design is to achieve a high level of modularity by partitioning the STM system into a set of easily separable, interchangeable components.  This naturally segregates the UHV needs of STM instrumentation from the typically non-UHV construction of a dilution refrigerator, facilitating the usage of non-UHV materials while maintaining a fully bakeable UHV chamber that houses the STM. The modular design also permits speedy removal of the microscope head from the rest of the system, allowing for repairs, modifications, and even replacement of the entire microscope head to be made at any time without warming the cryostat or compromising the vacuum. Without using cryogenic filters, we measured an electron temperature of $184$~mK on a superconducting Al($100$) single crystal.
\end{abstract}

\maketitle

\section{\label{sec:introduction}Introduction}

The diverse suite of scanning probe techniques available today are indispensable to modern condensed matter and materials research due to their unparalleled capabilities in providing spatially resolved measurements of an assortment of physical quantities. Invented in $1982$ by Gerd Binnig and Heinrich Rohrer, scanning tunneling microscopy (STM) was the first of these techniques to be developed, offering unprecedentedly high-resolution real-space imaging of material surface structures.\cite{binnig1982surface} Since then, a number of advances have been made to greatly expand the experimental versatility of STM, establishing the technique as a powerful spectroscopic tool for probing exotic electronic phenomena\cite{annurev2016} and as a highly sophisticated assembly tool for building nanoscale structures. 

One of the most important advancements in scanning tunneling spectroscopy (STS) was the integration of the dilution refrigerator within STM systems.\cite{moussy2001very, sueur2006room, kambara2007construction, song2010invited, marz2010, suderow2011, misra2013design, assig2013scanning, singh2013construction, roychowdhury2014scanning, allworden2018design, balashov2018compact, machida2018scanning} The implications of incorporating this cooling technique are two-fold. First, dilution-refrigerator STM systems have provided significantly higher energy resolution compared to ${}^3$He and helium-bath systems, which is crucial for differentiating the often-subtle spectroscopic features of quantum matter. This high energy resolution, for example, has been utilized to place an upper bound on the splitting of the zero-bias peak in Majorana atomic chains\cite{FeldmanDoubleEye}. Second, dilution-refrigerator technology has enabled the exploration of low-temperature electronic phases that only order at millikelvin temperatures and have been previously inaccessible to STM. Such exotic phenomena probed by dilution-refrigerator STM systems include d-wave superconductivity in CeCoIn${}_5$ \cite{Zhou2013}, many-body states in the quantum Hall regimes of graphene\cite{Song2010} and Bi(111)\cite{FeldmanNematic2016, Randeria2018, Randeria2019}, and zero-energy vortex bound states on Fe(Se,Te)\cite{Machida2019}.

Engineering systems with coexisting dilution refrigerators, which are typically non-UHV in construction, and STM instrumentation, which are ideally UHV in construction, is challenging due to the competing requirements of each individual subsystem. Researchers often give priority to the more pressing UHV needs of STM instrumentation, which results in often-inferior cryogenic wiring techniques and dilution-refrigerator constructions.
Moreover, achieving millikelvin temperatures requires thermally isolating the many stages of a dilution refrigerator from one another and the surrounding environment, which inherently increases the time necessary to warm an STM system for repairs and upgrades.

Here we describe the design and construction of an ultra-low-temperature scanning tunneling microscope (ULTSTM) that circumvents these issues with a modular design. Our system is neatly divided into five fully separable sub-systems: $1$) the liquid helium dewar and superconducting magnet, $2$) the dilution-refrigerator insert, $3$) the UHV chambers, $4$) the microscope head module, and $5$) the sample holders. We describe here the reasoning that has guided our design choices, the inter-connectivity of the various components, and the performance of the integrated system.

\begin{figure*}
\includegraphics[width=\linewidth]{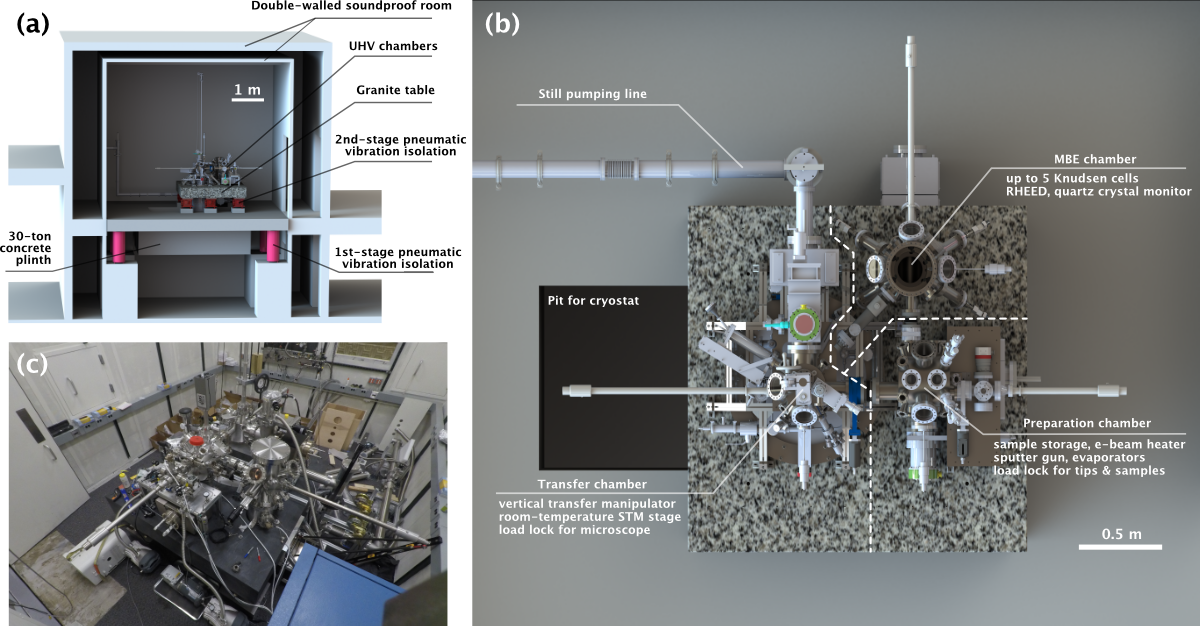}
\caption{\label{fig:room_layout} (a) The instrument is acoustically isolated from the outside environment by a double-walled soundproof room. The UHV system sits on a granite table on pneumatic isolators, which itself sits on a 30-ton concrete plinth on another set of pneumatic isolators. (b) The UHV system comprises three main chambers: the transfer chamber, preparation chamber, and MBE chamber. The transfer chamber contains a room-temperature microscopy stage and a vertical transfer manipulator for inserting the microscope head into the cryostat. The preparation chamber contains sample heaters, a sputter gun, and sample storage. (c) Photograph of the UHV system on the granite table.}
\end{figure*}

\section{\label{sec:vibration}Acoustic, Vibrational, and Electrical Isolation}

Since the tunneling between the tip and the sample is exponentially sensitive to their separation, the quality of topographic and spectroscopic data obtained using STM requires maintaining this distance with picometer-scale precision. Accomplishing such conditions requires reducing acoustic, vibrational, and electrical coupling between the microscope and its outside environment. The successful isolation of an STM facility from each of these sources of noise is a necessary condition for high-resolution imaging and local spectroscopy.

Acoustic isolation is achieved primarily through the use of a double-walled, soundproof instrumentation room, as depicted in Fig.~\ref{fig:room_layout}(a). We obtain further reduction of acoustic noise by tightly wrapping the dewar's sidewalls and bottom first with a one-eighth-inch-thick layer of $70$A neoprene rubber followed by a one-eighth-inch-thick layer of $30$A neoprene rubber. The rubber, which dampens the natural resonances of the dewar's sidewalls, is secured to the dewar by a set of two-inch-thick nylon straps (see Fig.~\ref{fig:dewar}(e)). Acoustic noise from the floor of the facility is blocked from reaching the STM system by a ``plinth plug" (made of wood and soundproof foam) that covers the bottom of the concrete plinth described in the next paragraph.

Vibrational isolation is achieved using a two-stage passive system consisting of a $2$-ton granite table sitting atop a $30$-ton concrete inertial plinth. Each mass rests on six pneumatic isolators with natural frequencies $7$~Hz and $1$~Hz that sharply attenuate vibrational noise from the laboratory floor. The UHV chambers are supported above the granite table while the cryostat is secured to a circular mounting plate on the top surface of the granite table. The cryostat is concentric with a circular hole cut into the granite table and subsequently passes through a rectangular hole in the center of the plinth. As mentioned above, this hole in the plinth is plugged by a wooden board to block acoustic noise from the facility floor from reaching the cryostat.

Additionally, special precautions were taken to reduce the transmission of acoustic and vibrational noise from the vacuum pumps needed to run the dilution-refrigeration system. These pumps reside in a room adjacent to the double-walled instrumentation room. For the $1$K-pot pumping line, flexible, stainless-steel vacuum bellows interface with the pumps and follow a three-meter-long, contorted path along the floor of the noisy room. The bellows are compressed by foam and heavy weights, which serve to scatter and dissipate acoustic and vibrational noise that travel along the length of the bellows.

The $1$K-pot pumping line and dilution-refrigerator still pumping line are both fed into a roughly one-quarter-cubic-meter concrete block in this separate noisy room. From there, the two vacuum lines are carried through the two walls of the instrumentation room into a set of two vibration-isolation gimbles. These prevent low frequency vibrations from transmitting into the room and ensure these lines do not mechanically short the concrete plinth to the outside environment. The vacuum lines then pass through a secondary concrete block of similar size sitting directly on the plinth inside the instrumentation room. Finally, the $1$K-pot pumping line is carried directly onto the granite table via a Kuri Tec plastic hose, while the still pumping line is passed through another isolation gimble before rigidly attaching to the cryostat.\cite{misra2013design}

\begin{figure}
\includegraphics[width=\linewidth]{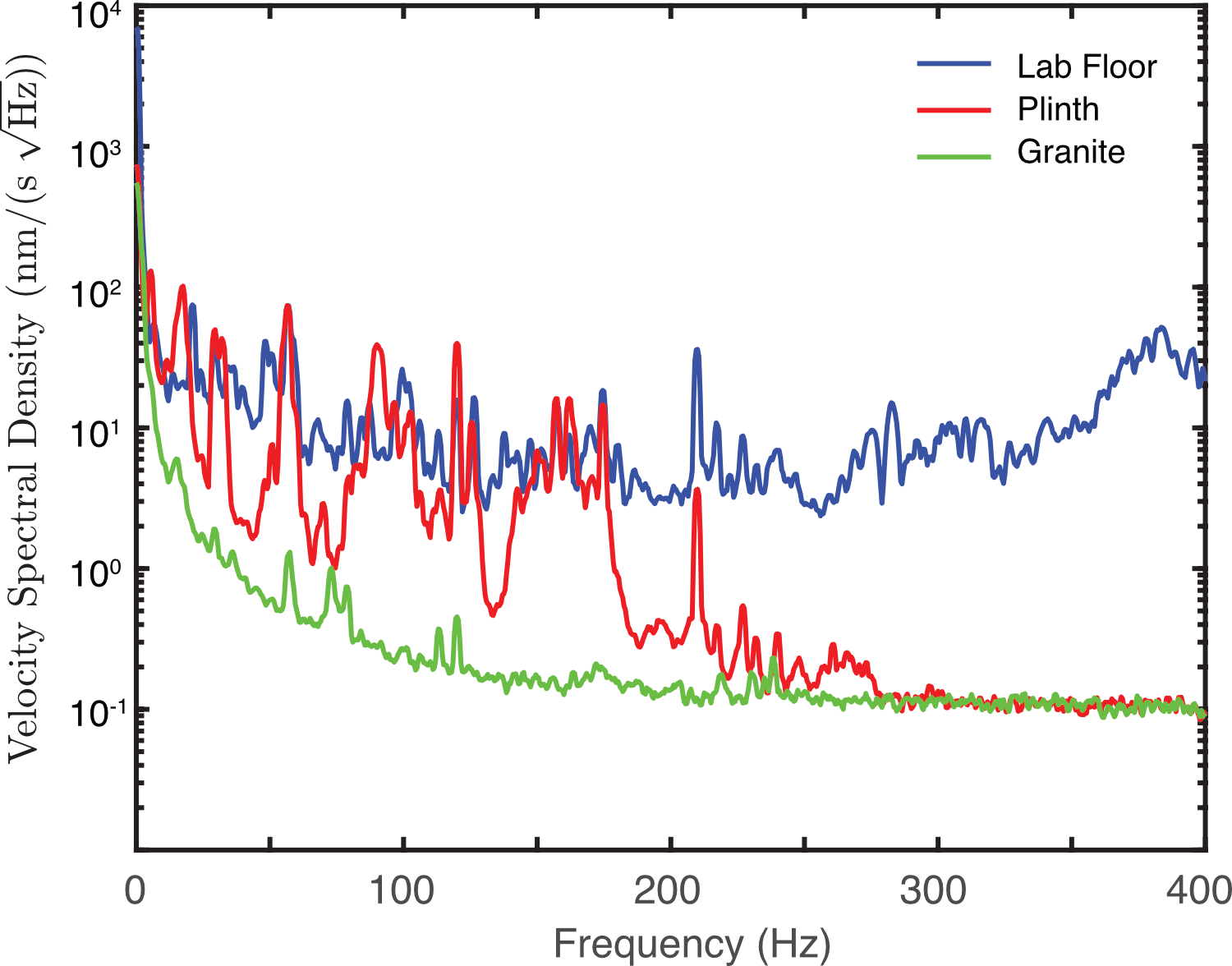}
\caption{\label{fig:vibrationspec} Velocity spectral density measured by a Wilcoxon $731$A accelerometer on the lab floor (blue), the concrete inertial plinth (red), and the granite instrumentation table (green). These data were acquired while running the dilution refrigerator.}
\end{figure}

Fig.~\ref{fig:vibrationspec} shows the velocity spectral density measured with an accelerometer (Wilcoxon $731$A) on the lab lower floor (blue), the concrete plinth (red), and the granite table (green) while the dilution refrigerator was running at base temperature. The velocity spectral density on the granite table is nearly commensurate with the noise floor of the accelerometer\cite{misra2013design}, with only a few additional peaks at various frequencies. These data demonstrate that each isolation stage dramatically reduces the transmission of external vibrational noise into our system.

Great care was taken to electrically isolate the microscope from external sources of electrical noise, such as the dilution-refrigerator pumps and the system's control electronics. All electrical grounds are connected to a single tether point on the transfer chamber's aluminum support frame by tinned copper braids with resistance on the order of a few milliohms or less, avoiding ground loops when possible. This grounding point is, in turn, tethered to a buried grounding plate by an additional thick, braided grounding cable. We have found that the grounding scheme has a large influence on the measured energy resolution of the microscope, with improper grounding and ground loops increasing the effective electron temperature measured on Al($100$) to as much as $400$~mK.

\section{\label{sec:dewar} Dewar and Superconducting Magnet}

The dewar, supplied by Oxford Instruments, has a $10$-day hold time and contains $210$ L of liquid helium, of which $180$~L is considered usable volume. The remaining liquid helium submerges a superconducting vector magnet that supports a $9$~T out-of-plane field and a $1$~T vector field in any direction. The $9$-$1$-$1$~T magnet has a bore diameter of $77$~mm. We are able to change the magnetic field at a rate of $0.01$~T/min while in tunneling, with no noticeable increase in noise. This could allow us perform measurements at a fixed spatial location while continuously varying the magnetic field.

\begin{figure*}
\includegraphics[width=\linewidth]{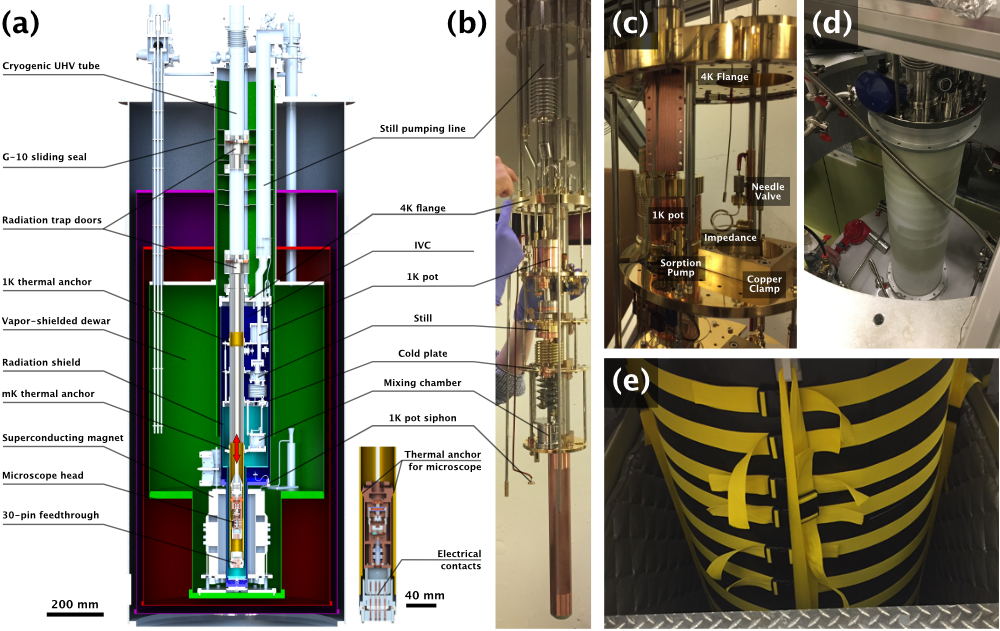}
\caption{\label{fig:dewar} (a) Schematic of the cryostat consisting of the dewar with the superconducting vector magnet, the dilution-refrigerator insert, and the UHV tube chamber. (b) Photograph of the insert with the UHV tube threading through a hole in each stage of the dilution refrigerator. The UHV tube is clamped to the insert via square-shaped copper thermalization blocks. (c) The $1$K-pot plate without the UHV tube. The needle valve and fixed impedance, which feed helium into the 1K pot, can be seen in the back. (d) Photograph of dropping the dewar from the insert. Since the cryostat is still filled with liquid helium, frost is building on the G-$10$ sliding seal (green tube surrounding the insert). (e) To dampen acoustic resonances in the dewar, we wrap the dewar in hard, black rubber tightly secured by yellow straps.}
\end{figure*}

The dewar is equipped with a G-$10$ sliding seal that allows the dewar to be attached to or detached from the dilution-refrigerator insert while liquid cryogens are still in the dewar (see Fig.~\ref{fig:dewar}(d)). This enables the ability to exchange the dewar-and-magnet system for a different dewar-and-magnet with different capabilities, without venting the UHV or warming the dilution-refrigerator insert to room temperature. In addition, this capability also enables us to warm a cold insert to room temperature within half a day for necessary repairs and desired modifications, removing the lengthy time required to warm the well-insulated dewar-and-magnet system.

\section{\label{sec:insert} Dilution refrigerator}

\subsection{\label{subsec:insert_overview} Overview and Design}

To achieve millikelvin temperatures for high-resolution spectroscopy, we use a dilution refrigerator (Fig. \ref{fig:dewar}(b)) designed and built in collaboration with Oxford Instruments. The dilution refrigerator has a cooling power of $350~\mu\textrm{W}$ at a mixing-chamber temperature of $100$~mK. The dilution-refrigerator insert is housed in the inner vacuum chamber (IVC), where the term ``inner" is used to differentiate this volume from the outer vacuum chamber (OVC) that surrounds and insulates the liquid helium bath. At the top of the IVC lies the refrigerator's $4$K flange, which is in physical contact with liquid helium in the main bath when the dewar is filled to capacity. Several thick, silver-coated copper braids run along the outside of the IVC from the bottom of IVC to the $4$K flange. These copper braids aid the cooling and thermalization of the entire insert as the liquid helium level in the dewar varies.

In conventional dilution-refrigerator units manufactured by Oxford Instruments\cite{misra2013design, machida2018scanning}, the $1$K pot is fed liquid helium through a siphon that runs outside the IVC and enters into the IVC volume at the $4$K flange. In our dilution refrigerator, the siphon enters through the bottom of the IVC and does not make contact with the $4$K flange. This is to ensure the helium in the siphon remains liquid, as the $4$K-flange temperature can be much greater than $4$~K when the liquid level in the dewar is low. We have found that the noise performance of our system is highly sensitive to the temperature and flow rate of the He into the $1$K pot, as we describe below. 

Our dilution refrigerator is separated into four temperature stages: the $1$K-pot plate, the still plate, the cold plate, and the mixing-chamber plate, which typically establish steady-state operating temperatures near $1.5$~K, $800$~mK, $280$~mK, and below $10$~mK, respectively. Mounted to the cold plate is a first-stage radiation shield that surrounds the mixing-chamber plate and the lowest section of the UHV tube. Surrounding this first-stage shield and the remainder of the dilution-refrigerator stages is the IVC's sidewalls, acting as a second-stage, $4$~K shield. Each stage is used to support and sequentially thermalize the UHV tube that passes through a two-inch-diameter hole in each plate. Copper clamps (Fig.~\ref{fig:dewar}(c)) at each temperature stage thermally connect the stage to the UHV tube.

Although the IVC surrounds the UHV tube, the vacuum in this volume is separated from the IVC volume (Fig.~\ref{fig:dewar}(a)). The microscope head, which is made of UHV-compatible components, resides in the UHV tube chamber. Meanwhile, UHV-incompatible materials, such as GE varnish, indium seals, metallic tapes, epoxies, and many other substances, can be employed inside the IVC without contaminating the UHV space. This gives us great flexibility in the design and construction of the dilution refrigerator, and also allows us to bake the UHV tube chamber while the insert is detached to achieve cleaner, lower-pressure environments for atomic-scale microscopy.

To facilitate the warming and cooling of the cryostat's insert, the IVC is equipped with a small volume of helium exchange gas. Mounted on the $1$K-pot plate is a charcoal-based sorption pump with a resistive heater that allows one to control the collection and release of this exchange gas. When the temperature of the sorption pump falls below $15$~K, the pump collects the exchange gas within the IVC, thermally isolating all stages of the dilution refrigerator from one another. When heat is locally applied to the sorbtion pump (between $50$ and $75$~mW), the exchange gas is released back into the IVC volume. A stainless-steel nut spacer sits between the pump and the $1$K-pot plate, effectively mitigating appreciable heating of the $1$K-pot plate when using the resistive heater.

This feature would be impossible in a conventional dilution-refrigerator STM design (where the dilution unit shares the UHV space with the microscope head) and greatly accelerates two processes. First, aided by this exchange gas, the insert can be warmed to room temperature in half a day when the dewar is detached from the IVC for necessary repairs or desired upgrades. Second, if the dewar containing liquid helium is attached to IVC, the exchange gas thermally connects all stages of the dilution refrigerator to the helium bath. This helps cool a warm microscope head, since the microscope head is thermally connected to the mixing chamber.

\subsection{\label{subsec:insert_wiring} Wiring}

To achieve reasonable base temperature, electrical wiring must be thermally anchored at every stage of the dilution refrigerator. Tip, sample, scanner, and capacitance-sensor lines start with uninsulated stainless-steel coaxial cables (New England Wire Technologies; similar to N$12$-$50$F-$257$-$0$, but without the insulating jacket) at room temperature. These wires then traverse the length of the fridge to the mixing-chamber plate, where they transition to uninsulated silver-coated copper coaxial cables (New England Wire Technologies; N$12$-$36$S+$00002$-$0$) before plugging into an electrical feedthrough at the bottom of the UHV tube. The cables are thermalized to the $4$K flange by running them through a washboard-shaped clamp.  Each cable is then thermalized to each of the remaining stages of the dilution refrigerator by wrapping $9$~inches of the cable around a copper bobbin mounted to each stage. The bobbins are coated with liberal amounts of GE varnish (IMI-$7031$), and we use Teflon tape to help hold the coaxial cables around the bobbins as the GE varnish dries. 

The walker lines are thermally anchored to each stage of the dilution refrigerator in a similar manner, except the walker lines consist of $0.01$-inch polyimide-insulated manganin wires (California Fine Wire) from room temperature to the $1$K-pot plate, superconducting NbTi coaxial cables from the $1$K-pot plate to the mixing-chamber plate, and silver-coated copper coaxial cables from the mixing-chamber plate to the electrical feedthrough on the UHV tube. To promote thermalization between the UHV tube and the mixing chamber, we secured the silver-coated copper coaxial cables to the walls of the UHV tube with easily removable copper tape.

We now consider the bandwidth of the coaxial cables, which is important for transmitting the tunneling current without loss and for attenuating high-frequency external noise that may degrade the electron temperature. The length of the coaxial cable from the STM tip to the top of the cryostat is about $4$~m, the core-to-shield capacitance is roughly $430$~pF, and the end-to-end resistance of the cable is around $200~\Omega$. Using the telegrapher's equations for a coaxial cable with $50~\Omega$ characteristic impedance, we estimate that the $-3$~dB bandwidth of the cable is $2.8$~MHz. This bandwidth is much larger than is required for conventional STM measurements and is sufficient for lossless transmission of the tunneling current.

\subsection{\label{subsec:insert_impedance} 1K Pot Operation}

The $1$K pot is fed liquid helium through a siphon that enters through the bottom of the IVC. Before the fluid enters the $1$K pot, however, the siphon branches into two paths; one path is controlled by a manual needle valve and can be completely shut off mechanically, while the other path is a fixed, capillary-tube impedance (Fig.~\ref{fig:dewar}(c)) made by feeding a "D-shaped" wire through a narrow, hollow tube. The impedance continuously supplies helium into the 1K pot. The needle valve is thermally anchored to the $1$K pot by several silver-coated copper braids\cite{RACCANELLI2001763}.

After completing the construction of our STM facility, we found that the helium flow rate through the impedance was too large, causing unmanageable vibrational noise in the microscope's tunneling current. To correct this, we mounted a $2$~k$\Omega$ nichrome-based resistor on the inlet of the impedance to regulate the flow of liquid helium into the $1$K pot\cite{lawes1kpot1998}. We also attached a Cernox thermometer on the needle valve to monitor and finely tune the power administered to the $2$~k$\Omega$ resistor.

The performance of our STM system strongly depends on the amount of heat applied to the impedance inlet in this flow-rate regulation procedure. When the heat applied to the impedance is insufficient, the $1$K pot begins to overfill and the tunneling current becomes noisy. When the heat applied to the impedance is excessive, the 1K pot empties and the ${}^3$He-${}^4$He mixture stops condensing. To maintain the $1$K pot level in a quiet, steady-state condition, we open the manual needle valve very slightly and supply roughly $50$~mW to the fixed impedance via the impedance heater. By making fine adjustments to the imparted power to maintain the needle-valve temperature at $3.9~\textrm{K} \pm 0.1~\textrm{K}$, we are able to run continuously. We discuss the noise performance of this setup in Sec.~\ref{subsec:copper_noise}.

\section{\label{sec:uhv}Ultra-High Vacuum Assembly}

\begin{figure*}
\includegraphics[width=\linewidth]{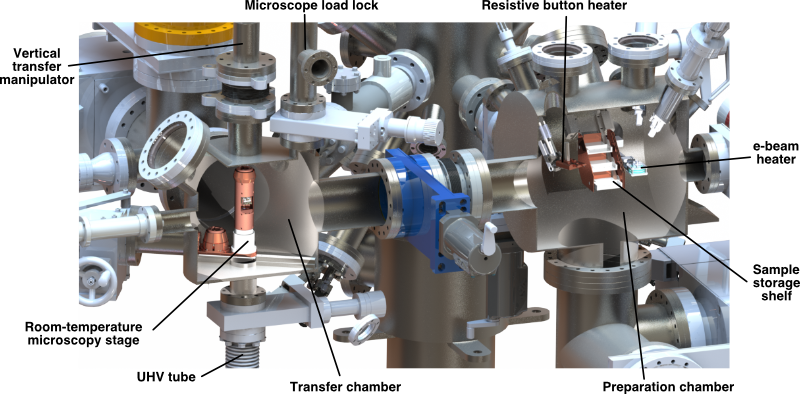}
\caption{\label{fig:chambers} Cross-sectional view of the transfer chamber and preparation chamber. The transfer chamber contains the room-temperature microscopy stage and is connected to the microscope load lock and the UHV tube chamber. The preparation chamber contains a resistive heater, an e-beam heater, and a sample storage shelf. An Ar${}^+$ sputter gun (not shown) is focused on the e-beam heater. The sample load lock (not shown) is connected to the right of the preparation chamber.
}
\end{figure*}

\subsection{\label{subsec:uhv_overview} Overview}

Depicted in Fig.~\ref{fig:room_layout}(b) are the three main UHV chambers: a transfer chamber, a preparation chamber, and an auxiliary chamber that is planned to be used for molecular-beam epitaxy (MBE). Attached to the preparation chamber is a sample load lock for introducing new samples and tips. Horizontal transfer manipulators and wobble sticks facilitate movement of components among the UHV chambers.

The preparation chamber (see Fig.~\ref{fig:chambers} for a cross-sectional view) provides a UHV space for {\em in situ} cleaning, analysis, and storage of samples. In the middle of the preparation chamber is a three-tier shelf upon which samples, tips, and miscellaneous gadgets (e.g. a wobble-stick-enabled Allen wrench) can be placed. On one side of the storage shelf is an electron-beam (e-beam) heater. The e-beam heater consists of two high-voltage vertical rods that constrain the lateral position of the sample holder (see Sec.~\ref{sec:sample_holders} for more details), which is placed above a thoriated-tungsten filament. An Ar${}^+$ sputter gun is aimed at the e-beam heater. On the other side of the storage shelf is a resistive heater (HeatWave Labs) and Type K thermocouple for annealing samples with more precise temperature control, but at lower temperatures than can be provided by the e-beam heater. A residual gas analyzer (Stanford Research Systems; RGA$100$) in the preparation chamber provides information about the quality of the vacuum and can be used to troubleshoot sample preparation procedures.

The transfer chamber (see Fig.~\ref{fig:chambers} for a cross-sectional view) serves three distinct purposes. First, it contains a room-temperature microscopy stage (see Sec.~\ref{subsec:rt_stage}) designed to connect to the microscope head via a $30$-pin military-style electrical feedthrough. This stage is used for accelerated sample diagnostics and for initially configuring the tip-sample positioning of micron-scale exfoliated devices using an optical-access viewport. Second, this chamber has an attached microscope load lock, which can be used to extract microscope modules for repair or to introduce microscope modules with differential capabilities without the need to compromise the vacuum or cryogenics of the facility. Finally, using a $2.7$-meter-tall vertical transfer manipulator (described in Sec.~\ref{subsec:vertical_manipulator}), the microscope head may be removed from the room-temperature stage and plunged into the cryostat-surrounded UHV tube chamber (see Sec.~\ref{subsec:uhv_tube}) hanging below the granite, plugging into an identically configured $30$-pin electrical feedthrough. Transferring the microscope from the room-temperature stage to the cryogenically cooled UHV tube chamber takes less than a minute.

\subsection{\label{subsec:rt_stage} Room-Temperature Microscopy Stage}

Within the transfer chamber is a room-temperature microscopy stage, which supports a $30$-pin electrical feedthrough used to secure and interface with the microscope head module. This allows us to perform measurements at room temperature, enabling us to quickly judge the quality of the samples before plunging the microscope head into the cryostat. The room-temperature stage is mounted on a rail (Fig.~\ref{fig:chambers}) that runs along the base of the transfer chamber and provides uniaxial translation.  Here, we describe three operational positions on this rail.

In the first position, the $30$-pin electrical feedthrough sits directly beneath the vertical transfer manipulator and above the UHV tube chamber. At this position, samples and STM tips can be inserted into or removed from the microscope head using a wobble stick. The samples are in line of sight of three $2.75$" viewports that provide optical access for tip-sample alignment and can be outfitted with evaporators for atomic and molecular deposition. In this position, the microscope head can also be grabbed by the vertical transfer manipulator through a mechanism fully described in Sec.~\ref{subsec:vertical_manipulator}.

In the second position, the $30$-pin electrical feedthrough is directly beneath a $3.375$" gate valve that leads to an auxiliary vacuum chamber. This chamber serves as a microscope load lock and holds a shorter, $0.7$-meter-tall vertical transfer manipulator that allows us to extract the microscope head from the transfer chamber. We can then repair and upgrade the capabilities of the microscope or replace the microscope head without venting the transfer chamber or warming the cryostat.

In the third position, the room-temperature microscopy stage is fully retracted, giving clearance for the vertical transfer manipulator to be inserted into the UHV tube chamber. It is in this position that the microscope head can be plugged into a $30$-pin electrical feedthrough at the bottom of the UHV tube chamber, within the cryostat.

In addition to the aforementioned features, the room-temperature microscopy stage can optionally be cooled with liquid nitrogen, effectively precooling the microscope head before insertion into the cryostat. This may be useful for keeping the microscope head at a low temperature for adatom deposition.

\subsection{\label{subsec:uhv_tube} Ultra-High Vacuum Tube}

\begin{figure}
\includegraphics[width=\linewidth]{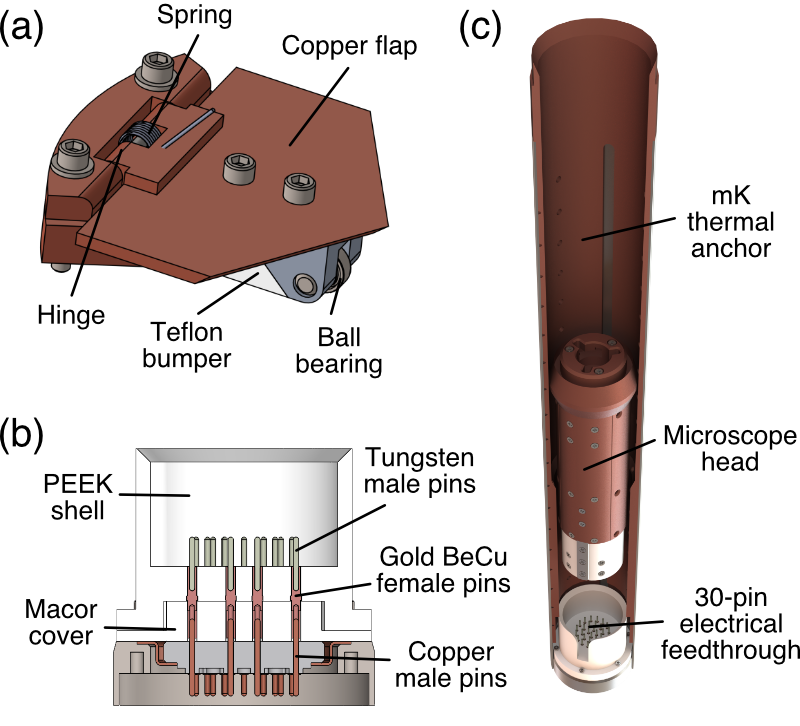}
\caption{\label{fig:uhv_tube_items} (a) Inside the UHV tube chamber are two copper trap doors that block radiation from reaching the microscope head at the bottom of the UHV tube chamber. Each trap door consists of three spring-loaded hinges that automatically close three, symmetric copper flaps. Mounted on the top face of every copper flap is a Teflon bumper and ball bearing that makes contact with the microscope head and $1$K pot plug when they are lowered into the UHV tube chamber via the vertical transfer manipulator. The bumpers and ball bearings ensure these items are not caught on the triangular corners of the copper flaps. (b) Cross-sectional view of the $30$-pin electrical feedthrough welded to the bottom of the UHV tube. The room-temperature microscopy stage has a similar $30$-pin feedthrough. (c) Cartoon of the microscope head being lowered into the UHV tube chamber. A guiding key on the microscope head aligns with a slot cut into the PEEK shell of the $30$-pin feedthrough. The top of the microscope head makes contact with a constriction in the mK thermal anchor.
}
\end{figure}

The UHV tube is a $2$-meter-long, $2$"-diameter tube (with $0.035$"-thick, stainless-steel walls) that extends into the cryostat. At the top of the UHV tube is a $3.375$" gate valve that isolates the UHV tube chamber from the transfer chamber when the transfer chamber needs to be vented. At the bottom of the UHV tube chamber (which is surrounded by the magnet) is a $30$-pin electrical feedthrough identical to that which is on the room-temperature microscopy stage. The $30$-pin feedthrough (custom part from Cosmotec) is schematically drawn in Fig.~\ref{fig:uhv_tube_items}(b). The microscope head plugs into the $30$-pin feedthrough, electrically connecting the samples, tip, scanner, walkers, thermometers, and capacitance sensors to the silver-coated copper wires mentioned in Sec.~\ref{subsec:insert_wiring}.

In the upper half of the UHV tube chamber, above the dilution-refrigerator stages, are two radiation trap doors (see Fig.~\ref{fig:dewar}(a)). Each trap door consists of three spring-loaded copper flaps (see Fig.~\ref{fig:uhv_tube_items}(a)) that block the line of sight from the transfer chamber into the UHV tube chamber and, thus, prevent radiation from reaching the microscope. The trap doors open manually when the vertical transfer manipulator is lowered into the UHV tube chamber. Ball bearings and Teflon bumpers prevent the vertical transfer manipulator and the bottom of the microscope from being caught on the trap doors.

The walls of the UHV tube are mostly stainless steel, with the exception of a set of two, in-line inserted copper ring sections that are used to thermalize the UHV tube at the $1$K-pot and mixing-chamber stages (see Fig.~\ref{fig:dewar}(a)). The first copper section (henceforth referred to as the 1K thermal anchor) is large enough to pass the microscope head, but is designed with a radial, conical constriction that matches the outer dimensions of a copper radiation block (which we refer to as the $1$K pot plug) that further prevents radiation from reaching lower parts of the UHV tube chamber. The top of the $1$K pot plug is identical in design to the top of the microscope head, allowing it to interface with the vertical transfer manipulator in a similar fashion. The bottom of the $1$K pot plug is shaped such that the plug can be inserted into a $30$-pin electrical feedthrough, which allows the plug to be carried by the rail on the room-temperature microscopy stage to the microscope load lock. The $1$K pot plug is equipped with three spring-loaded latches that mate with three slotted grooves in the 1K thermal anchor, preventing the $1$K pot plug from rotating when engaged with the vertical transfer manipulator.

The second copper section (henceforth referred to as the mK thermal anchor) contacts the mixing-chamber plate and runs along the lower length of the UHV tube chamber. The mK thermal anchor is designed with an additional radial, conical constriction which seats the top section of the microscope, providing a direct thermal connection to the mixing chamber when the base of the microscope is plugged into the UHV tube chamber's $30$-pin electrical feedthrough (see Fig.~\ref{fig:uhv_tube_items}(c)).

The UHV tube is clamped at each stage of the dilution-refrigerator insert by a set of fully removable copper clamps. Loosening these clamps allows the insert to be detached from the UHV tube so that the UHV tube can be baked.

\subsection{\label{subsec:vertical_manipulator} Vertical Transfer Manipulator}

\begin{figure}
\includegraphics[width=\linewidth]{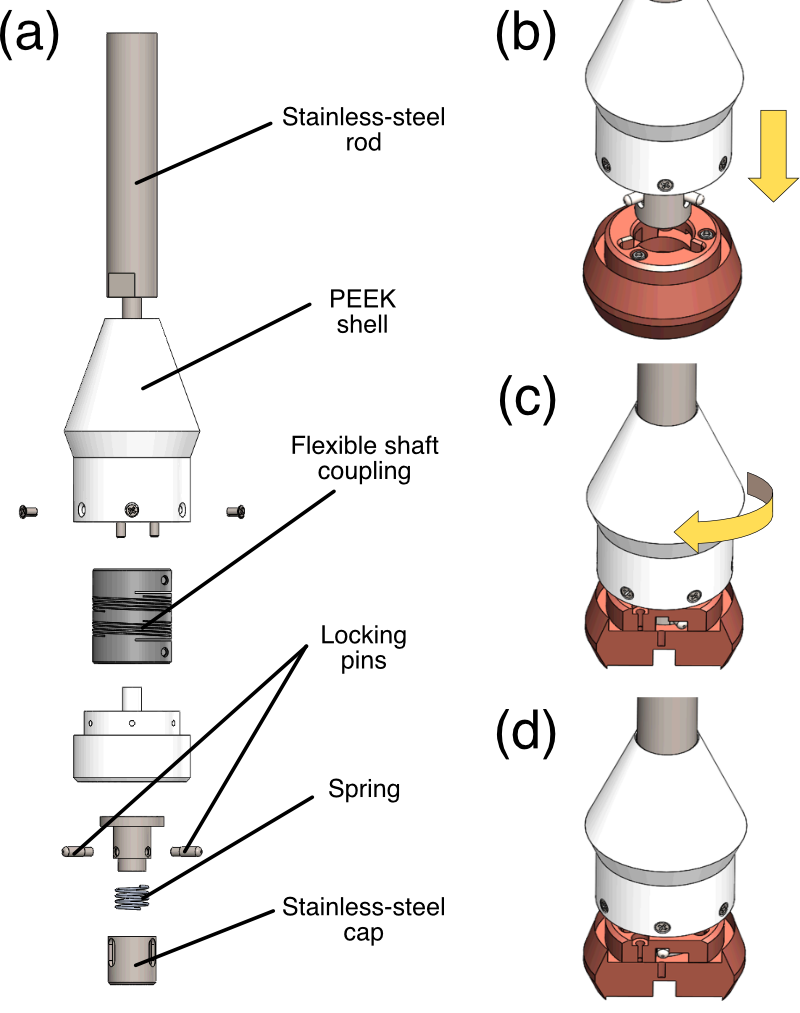}
\caption{\label{fig:locking_mech} (a) Exploded-view drawing of the bayonet-mount-style locking mechanism at the end of the vertical transfer manipulator in the transfer chamber and in the microscope load lock. This locking mechanism interfaces with the top of the microscope head and $1$K pot plug to move them between the microscope load lock, the room-temperature microscopy stage, and into the cryostat-surrounded UHV tube chamber. (b) Three locking pins align with and enter three slots on the top of the microscope head. (c) The locking pins are inside J-shaped grooves cut into the top of the microscope head. (d) Rotating the locking mechanism from the position in (c) brings the locking pins into the serifs at the end of the grooves and fastens the locking mechanism to the microscope head.
}
\end{figure}

A magnetically coupled vertical transfer manipulator (UHV Design) enables the ability to pull the microscope head and $1$K pot plug from the room-temperature microscopy stage and insert these items into the UHV tube chamber, and vice versa. Translational motion in the vertical direction is achieved through a pulley with a brass counterweight, while rotation motion is achieved using a ball spline. Three stainless-steel cables are anchored to the granite table and extend to the top of the vertical transfer manipulator to dampen oscillations and help align the translation axis of the vertical transfer manipulator to the UHV tube.

In order to shuttle the microscope head or $1$K pot plug between the room-temperature microscopy stage and the UHV tube chamber, the vertical transfer manipulator is equipped with a locking mechanism akin to a bayonet mount, as depicted in Fig.~\ref{fig:locking_mech}. Three radially pointing pins interlock with a set of three J-shaped grooves cut into the top of the microscope. These radial pins are attached below a flexible shaft coupling to ensure alignment between the pins and the microscope head.

\section{\label{sec:stmhead} Microscope Head}

\begin{figure*}
\includegraphics[width=\linewidth]{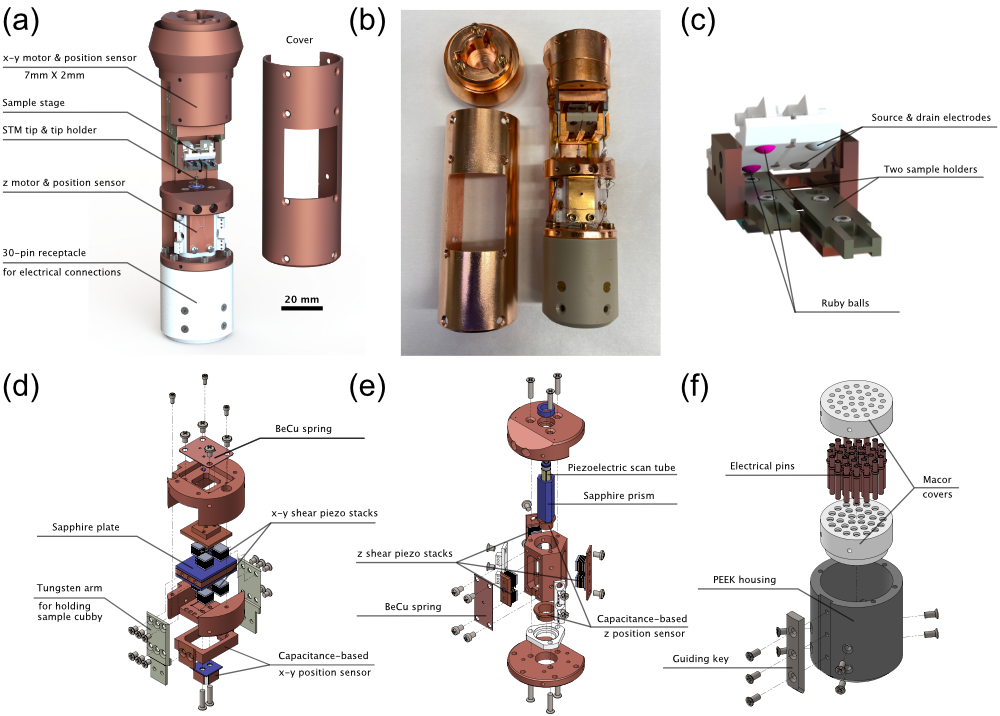}
\caption{\label{fig:microscope_head} (a) Schematic of microscope head. At the top is a locking mechanism that interfaces with the vertical transfer manipulator, followed by an x-y motor that provides lateral positioning of the samples. The x-y motor is coupled via tungsten arms to a sample cubby that is capable of holding two sample holders. Underneath the sample cubby is a Pan-style walker that moves the tip and tube scanner in the vertical direction. At the bottom of the microscope head is a receptacle with $30$ electrical pins. (b) Photograph of microscope head and cover. (c) Schematic of the sample cubby. When the sample cubby is closed, ruby balls press against and secure the left sample holder, while tungsten balls secure the right sample holder. The tungsten balls also act as auxiliary source and drain electrodes for samples that support multiple-terminal measurements. (d) Exploded-view drawing of the x-y walker. (e) Exploded-view drawing of the z walker. (f) Exploded-view drawing of the $30$-pin receptacle.}
\end{figure*}

The microscope head is detailed in Fig.~\ref{fig:microscope_head}(a). At the top of the microscope head is a bayonet-mount-style fastening mechanism that interfaces with the vertical transfer manipulator. Below that is an x-y coarse walker (Fig.~\ref{fig:microscope_head}(d)) that consists of a polished sapphire plate (Applied Ceramics) sandwiched between six shear piezo stacks (Physik Instrumente). The six piezo stacks are each made of interleaved piezoelectric plates poled in orthogonal directions, which provide stick-slip actuation of the sapphire plate in both the x- and y-directions, with a full range of motion of roughly $7$~mm and $2$~mm, respectively. This large range of motion is crucial for the double-sample-holder cubby design, described below, and effective navigation to micron-size devices. Below the x-y walker is a $4$-contact capacitance sensor consisting of a rectangular copper ring that moves with the sapphire plate and four fixed BeCu plates (see Fig.~\ref{fig:microscope_head}(d)). Measuring the response on the BeCu plates to an a.c. voltage applied on the copper ring provides information about the lateral position of the sapphire plate with a resolution of about $100~\mu \mathrm{m}$.

The x-y walker is directly coupled to a sample cubby by vertical tungsten arms. The sample cubby, depicted in Fig.~\ref{fig:microscope_head}(c), is capable of holding two sample holders (described in more detail in Sec.~\ref{sec:sample_holders}), and the x-y walker is capable of moving the cubby such that the STM tip can approach the centers of each sample holder. The left sample holder is intended to hold a metallic crystal, such as Cu($111$) or Au($111$), for preparing the STM tip. The right sample holder is intended to hold the sample of interest. Sample holders are easily inserted into the sample cubby's two slots when a polyether ether ketone (PEEK) lid is opened using a wobble stick. When the PEEK lid is closed, BeCu springs push against two ruby balls over the left sample holder and two tungsten balls over the right sample holder.  These balls press against screws in counterbored holes in the backs of the sample holders, rigidly securing the sample holders to the cubby. The sample cubby is also thermally anchored to the body of the microscope head via flexible silver braids.

In addition to mechanically fastening the sample holders to the cubby, the tungsten balls also serve as electrical contacts to the right sample holder. While the body of the sample cubby acts as the electrode for sample bias for the left sample holder, the right sample holder supports three electrical connections. For samples that are in a field-effect-transistor (FET) geometry, the body of the sample cubby acts as the gate electrode, while the tungsten balls act as contacts for the source and drain electrodes.

The STM tip can be replaced {\it in situ} by using a wobble stick to plug the tip into a BeCu collet. This collet is epoxied into a piezoelectric tube (EBL Products Inc.; EBL $\#4$) that provides about $3~\mu \mathrm{m}$ lateral motion at room temperature and about $1~\mu \mathrm{m}$ lateral motion at base temperature. The piezoelectric tube is epoxied into a Macor support structure, which in turn is epoxied inside the cylindrical cavity of a sapphire prism (Applied Ceramics).

The sapphire prism is part of a Pan-style coarse walking mechanism \cite{panwalker1999} (Fig.~\ref{fig:microscope_head}(e)) that moves the STM tip in the z-direction, varying the tip-sample distance. Three highly polished faces of the sapphire prism are each adjacent to pairs of shear piezo stacks (Physik Instrumente) poled in the vertical direction. The force between the piezo stacks and the sapphire interface is controlled by compressing a BeCu leaf spring against a ruby ball, which presses against a copper plate upon which two of the piezo stacks are epoxied.

The bottom of the microscope head is a receptacle (Fig.~\ref{fig:microscope_head}(f)) that mates with the electrical feedthroughs on the room-temperature microscopy stage and at the bottom of the UHV tube chamber. The receptacle consists of $30$ copper pins (Solid Sealing Technology; KT$13353$-$11$) sandwiched between two Macor covers, all encased in a PEEK housing. A parallel key mounted on the back of the PEEK housing guides and aligns the insertion of the microscope head into the electrical feedthroughs (see Fig.~\ref{fig:uhv_tube_items}(c)).

Since the $30$ pins in the receptacle are not coaxial, neighboring pins have non-negligible crosstalk between them. It was crucial to the noise performance of the microscope to arrange these pins in a layout that best reduces interference between the tunneling current signal and other voltage signals, such as the scanner and walker lines. To achieve this, the pins for the tunneling current and the scanner and walker lines are placed on opposite sides of the receptacle. Furthermore, all pins surrounding the tunneling current pin are grounded during normal operation, forming a partial GHz-range Faraday cage around the tunneling current pin.

Two thermometers, a calibrated RuO$_2$ sensor (Lake Shore Cryotronics, Inc.; RX-$102$A) and a Cernox (Lake Shore Cryotronics, Inc.; CX-1010), were installed on the copper body surrounding the scanner. Each sensor covers a different temperature range; the RuO$_2$ covers $5$~mK to $40$~K, while the Cernox covers $1.2$~K to $330$~K.

The microscope head is also surrounded by a copper cover that protects the microscope. This cover has a window cut-out that gives the wobble stick direct access to the tip and sample. The various parts of the microscope head (including the cover) are fastened together using non-magnetic silicon aluminum bronze screws (Swiss Screw Products, Inc.).

\section{\label{sec:sample_holders} Sample Holders}

\begin{figure}
\includegraphics[width=\linewidth]{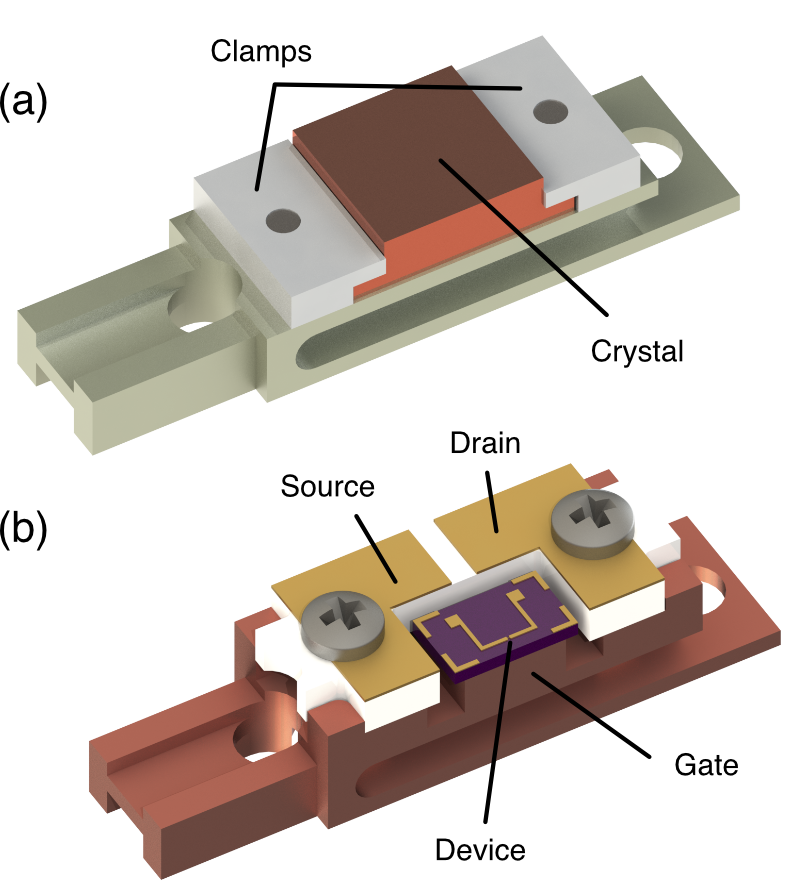}
\caption{\label{fig:sample_holders} (a) Tungsten sample holder for single crystals. The sample holder has two holes for aligning the sample on an e-beam heater such that the crystal sits at the focal point of an Ar${}^+$ sputter gun. During STM measurements, sample bias is applied to the body of the sample holder. (b) Copper sample holder for FET devices. A gate voltage is applied to the body of the sample holder, while source and drain voltages are applied to two pads that are insulated from the sample-holder body via Macor. The two pads are either gold evaporated onto the Macor or are copper foil, and wires connecting to the device are either glued or wirebonded to the pads.
}
\end{figure}

Fig.~\ref{fig:sample_holders} depicts two types of sample holders that are compatible with the sample cubby described in Sec.~\ref{sec:stmhead}. The first design (Fig.~\ref{fig:sample_holders}(a)), made of tungsten, is used for mounting single crystals and cleavable samples. Tungsten is required here for high-temperature annealing procedures. The second design (Fig.~\ref{fig:sample_holders}(b)), made of copper, is for gate-tunable device samples. Copper is used here for its superior thermal conductivity. The main body of the sample holder serves as the sample-bias electrode for a single-crystal sample holder and the gate electrode for a device sample holder.

Both varieties of sample holders share a common base structure consisting of two $0.1$" through holes and two guiding side rails. The holes are used to stabilize and center the sample holders on the e-beam heater in the preparation chamber within the line of sight of the Ar${}^+$ sputter gun. The side rails guide the sample holders into either side of the microscope head's sample cubby.

Between the aforementioned $0.1$" through holes are two additional counterbored holes for screws used for sample mounting. For a single-crystal sample holder, the screws fasten two overhanging clamps that secure the single crystal to the sample holder. For a device sample holder, the screws secure a Macor plate that electrically isolates the source and drain electrodes from the rest of the sample holder, which serves as the device back-gate. The screws in a device sample holder also electrically connect the source and drain electrodes on the sample holder to the tungsten balls in the sample cubby (see Sec.~\ref{sec:stmhead}), upon which source and drain bias voltages are applied.

\section{\label{sec:performance}Microscope Operation and Performance}

\subsection{\label{subsec:operation} Tip and Sample Exchange}

The modularity of the microscope is leveraged for the tip and sample exchange process. By interlocking the vertical transfer manipulator and the top of the microscope head, the cold microscope head is pulled from the 30-pin electrical feedthrough at the bottom of the UHV tube and then plugged into the equivalent 30-pin feedthrough on the room-temperature microscopy stage. From that position, a wobble stick has direct access to exchange the STM tip and samples. The microscope head is then pulled from the room-temperature microscopy stage via the vertical transfer manipulator and reinserted into the UHV tube. This process typically warms the microscope head to about $40$ to $60$~K, depending on the amount of time the microscope head spends holstered in the room-temperature microscopy stage.

From this elevated temperature, the microscope is cooled by physical contact with the mK thermal anchor in the UHV tube, which is thermally connected to the liquid helium bath through the controlled release of helium exchange gas from the sorption pump in the IVC. The microscope cools from $60$~K to $6$~K in about $3$ hours. From there, we can start the circulation of ${}^3$He-${}^4$He mixture, which requires an additional $3$ hours to cool from $6$~K to base temperature. In practice, we often wait until the next morning before performing measurements because the thermal contractions and increased liquid helium boil-off caused by this cooling process cause significant noise and require a few additional hours of settling time.

\subsection{\label{subsec:copper_noise} Noise Level on Cu(111)}

\begin{figure}
\includegraphics[width=\linewidth]{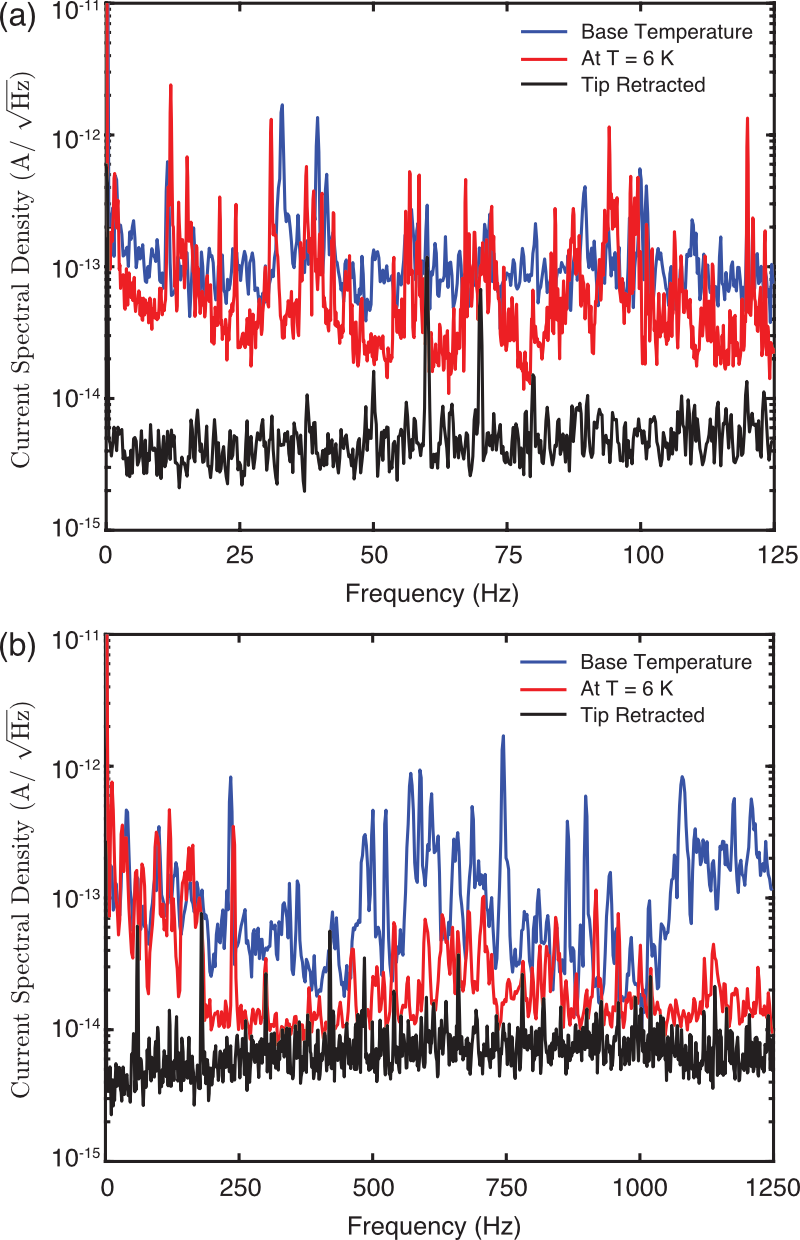}
\caption{\label{fig:currentspec} Open-feedback current spectral density over a frequency range below (a) $125$~Hz and (b) $1250$~Hz on Cu($111$) at $V_s = -200~\textrm{mV}$ and $I = 100~\textrm{pA}$ while the dilution refrigerator was not running (red curves; $T = 6~\textrm{K}$) and while the dilution refrigerator was running (blue curves; base temperature). The black curves were acquired when the tip was withdrawn from the surface ($I = 0~\textrm{pA}$).
}
\end{figure}

Our instrument provides a similar level of noise performance to that of other dilution-refrigerator STMs \cite{misra2013design, machida2018scanning}. The data presented here were obtained using a Femto DLPCA-$200$ current preamplifier at a gain of $10^9$~V/A and a bandwidth of $1$~kHz. The tunneling current signal from the preamplifier was fed into a $5$th-order Butterworth LC low-pass filter with a cutoff frequency of $1.1$~kHz and recorded by a Nanonis SPM Controller (SPECS GmbH). Simultaneously, the unfiltered tunneling current signal was also provided as input into a Stanford Research Systems SR$830$ lock-in amplifier to yield the tunneling differential conductance ($dI/dV$). The d.c. component of the sample bias and the scanner lines were also filtered by $1$~kHz LC low-pass filters located near the Nanonis SPM controller. Control over the flow rate of liquid helium into the $1$K pot (see Sec.~\ref{subsec:insert_impedance}) was essential for achieving this noise level.

Fig.~\ref{fig:currentspec} shows current spectral density measured with a PtIr tip on a Cu($111$) surface prepared by a standard sputter-and-anneal procedure. These data were acquired with an open-feedback tunneling setpoint of $V_s = -200~\textrm{mV}$ and $I = 100~\textrm{pA}$, where $V_s$ is the voltage applied to the sample and $I$ is the d.c. current measured from the tip. Plotted in black are the data acquired with the tip retracted from the sample. Plotted in red are the data acquired with the tip in tunneling at $T = 6~\textrm{K}$, while the dilution refrigerator was not running and all pumps were shut off. Plotted in blue are the data acquired with the tip in tunneling at base temperature, while the dilution refrigerator was in full, continuous operation. The magnetic field was zero during these measurements.

Some of the peaks in Fig.~\ref{fig:currentspec} can be directly attributed to particular residual sources of noise. Integer multiples of $60$~Hz are electrical, while peaks near $81$~Hz, $85$~Hz, and $243$~Hz can be attributed to natural resonances in the helium dewar's sidewalls. The dewar wall resonances are greatly attenuated by the rubber dewar wrapping scheme described in Sec.~\ref{sec:vibration}, but they are still detectable in the current spectral density. Also, differences in the current spectral density between $T = 6~\textrm{K}$ (red curves) and base temperature (blue curves) are largely due to vibrational noise induced by the flow of liquid helium into the 1K pot. We can reduce this noise by heating the capillary-tube impedance that feeds the 1K pot (see Sec.~\ref{subsec:insert_impedance}).

\subsection{\label{subsec:aluminum_gap} Superconducting Gap on Al(100)}

Eliminating ground loops and ensuring that ground path resistances are on the order of a few milliohms or less was essential for reducing the electron temperature and obtaining high energy resolution. Additionally, we ground the capacitance sensors, thermometers, and walkers during measurements at base temperature.

It is also crucial to attenuate radio-frequency (RF) noise that may excite electrons at the tunnel junction and raise the electron temperature. Thus, we placed a room-temperature RF filter (Mini-Circuits; VLFX-$80$) and a voltage divider on the bias line as close as possible to the cryostat. We also placed $\pi$-filters (API Technologies; $1204$-$050$) with $8$~dB insertion loss at $10$~MHz and $70$~dB insertion loss at $10$~GHz on the scanner lines. The tunneling current line between the microscope and the preamplifier remains unfiltered (except by the native capacitance and resistance of coaxial wiring in the insert) because a large input capacitance into the preamplifier leads to unstable current-to-voltage conversion.

\begin{figure}
\includegraphics[width=\linewidth]{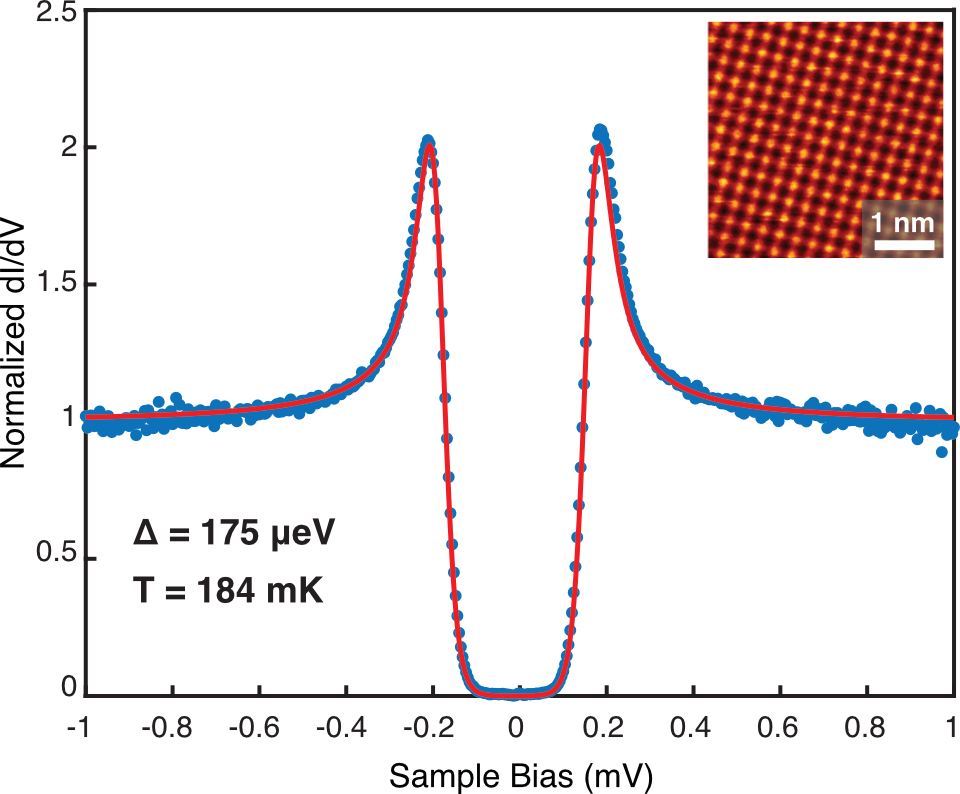}
\caption{\label{fig:alum} Differential conductance (blue dots) acquired on a superconducting Al($100$) surface, along with a fit to the Dynes function (red curve). The data is normalized such that the Dynes function is unity at infinity. The fit yields a superconducting gap of $\Delta = 175~\mu\textrm{eV}$ and an effective electron temperature of $184 \pm 7~\textrm{mK}$. Initial tunneling parameters: $V_s = -1~\textrm{mV}$, $I = 1~\textrm{nA}$, $4~\mu\textrm{V}$ a.c. modulation at $381.7~\textrm{Hz}$. The inset shows an atomically resolved image of clean Al($100$) with $V_s = -1~\textrm{mV}$ and $I = 5~\textrm{nA}$.
}
\end{figure}

Fig.~\ref{fig:alum} shows $dI/dV$ acquired on clean Al($100$) using a PtIr tip, prepared on Cu($111$), at zero magnetic field. Although the mixing chamber was at $15$~mK, the temperature measured by the RuO$_2$ thermometer on the microscope head was $110$~mK, indicating that there is still room for improving the thermal connection between the microscope and the mixing chamber. A fit to the Dynes function\cite{dynes1978} yields a Bardeen-Cooper-Schrieffer (BCS) quasiparticle gap of $\Delta = 175~\mu\textrm{V}$ (consistent with previous reports\cite{wells1970superconducting}) and an effective electron temperature of $184 \pm 7~\textrm{mK}$. Sec.~\ref{sec:future} discusses our strategy for improving the energy resolution in the future.

\section{\label{sec:future}Outlook and Future}

With this instrument, we recently probed strong correlations in magic-angle twisted bilayer graphene (MATBG)\cite{cascades} at $T = 6~\textrm{K}$, well above its superconducting transition temperature. Since superconductivity in MATBG is expected to be unconventional\cite{Cao2018}, we will extend our measurements to the millikelvin-temperature regime to study the symmetry of its order parameter. Two-dimensional (2D) materials offer a wide variety of interesting correlated phenomena amenable to investigation with spatially resolved techniques, and looking forward, there are a variety of improvements that would make our instrument a more effective tool for studying such phenomena.

The feasibility of modifications and upgrades to our instrument using an iterative engineering strategy is greatly enhanced by the modular design of our facility. For example, as of the writing of this manuscript, we are in the process of assembling a second microscope head module with key improvements over the original microscope head described in Sec.~\ref{sec:stmhead}. When the construction of this second microscope is completed, we can simply replace the original microscope head with the new microscope head via extraction through the microscope load lock. The microscope exchange process of warming the original module to room temperature, baking the new module in the microscope load lock, and cooling this module to cryogenic temperatures will only take two days and will not require venting the UHV system or warming the dewar.

This second-generation microscope head was designed with adjustments to its physical structure to ensure better thermal contact to the mK thermal anchor in the UHV tube chamber. It has also been designed with a set of six silver-epoxy cryogenic filters\cite{scheller2014filter} to reduce the electron temperature. A future version of the microscope head may also include additional contacts to support four-terminal electrical transport measurements as well as non-contact atomic force microscopy (nc-AFM)\cite{gross2009}. Our modular design enables rapid testing of these alterations with only minimal disruptions to ongoing experiments.

Since the dewar can slide freely against the insert, we can quickly swap dewars without compromising the UHV or warming the insert to room temperature. The dewar currently attached to our system is equipped with a vector magnet that provides $9$~T in the out-of-plane direction or $1$~T in any direction. We are working with Oxford Instruments to obtain an alternate dewar that can supply a significantly larger out-of-plane field. A dewar with a $16$~T magnet would have the exact same dimensions as our current $9$-$1$-$1$~T magnet, but modifications to the facility structure can be made to support a magnet with a field of $18$~T or larger.

\begin{acknowledgments}

For helpful discussions and/or technical assistance, the authors thank Steve Shedd, Peter Wilson, and Ulrich Marckmann from Integrated Dynamics Engineering, Nick Dent and John Burgoyne from Oxford Instruments, Steven Lowe and William Dix from the Princeton Department of Physics machine shop, Joseph A. Stroscio from the National Institute of Standards and Technology (NIST), and Nana Shumiya, Berthold J\"{a}ck, Hao Ding, Mallika Randeria, Yonglong Xie, and Hiroyuki Inoue from Princeton University. The instrumentation and infrastructure were primarily supported by the Gordon and Betty Moore Foundation, with additional funding from NSF-DMR-1608848, NSF-DMR-1904442, DOE-BES grant DE-FG02-07ER46419, ONR  
N00014-17-1-2784, the Eric and Wendy Schmidt Transformative Technology Fund at Princeton, and the NSF-MRSEC program through the Princeton Center for Complex Materials DMR-1420541.

This article may be downloaded for personal use only. Any other use requires prior permission of the author and AIP Publishing. This article appeared in {\em Rev. Sci. Instrum.} {\bf 91}, 023703 (2020) and may be found at \url{https://doi.org/10.1063/1.5132872}.

\end{acknowledgments}

\medskip


\bibliography{bibliography}

\end{document}